\documentclass[11pt,draftcls,onecolumn]{article}
\usepackage{multicol,caption}
\usepackage{multirow}
\usepackage[a4paper, total={7in, 9in}]{geometry}
\usepackage{graphicx}
\usepackage{adjustbox}
\usepackage{changepage}
\usepackage{enumitem}
\usepackage{listings}
\usepackage{amsmath}
\usepackage{color}
\usepackage{rotating}
\usepackage{pdfpages}
\usepackage{amsfonts}
\usepackage{tikz}
\tikzset{
	treenode/.style = {shape=rectangle, rounded corners,
		draw, align=center,
		top color=white, bottom color=blue!20},
	root/.style     = {treenode, font=\Large, bottom color=red!30},
	env/.style      = {treenode, font=\ttfamily\normalsize},
	dummy/.style    = {circle,draw}
}

\usepackage{listings}
\usepackage{courier}

\usepackage{natbib}
\usepackage{apalike}

\newenvironment{Figure}
{\par\medskip\noindent\minipage{\linewidth}}
{\endminipage\par\medskip}

\newcommand{\bc}{\begin{center}}
\newcommand{\ec}{\end{center}}

\newtheorem{defn}{Definition}
\newtheorem{thm}{{\cal T}heorem}[section]

\newtheorem{prop}{Proposition}
\newtheorem{lem}[thm]{Lemma}
\newtheorem{conj}[thm]{Conjecture}
\newtheorem{constr}[thm]{Construction}
\newtheorem{note}{Remark}
\newcommand{\bdefn}{\begin{defn}}
	\newcommand{\edefn}{\end{defn}}
\newcommand{\bnote}{\begin{note}}
	\newcommand{\enote}{\end{note}}
\newcommand{\bprop}{\begin{prop}}
	\newcommand{\eprop}{\end{prop}}
\newcommand{\blem}{\begin{lem}}
	\newcommand{\elem}{\end{lem}}
\newcommand{\bthm}{\begin{thm}}
	\newcommand{\ethm}{\end{thm}}
\newcommand{\bconj}{\begin{conj}}
	\newcommand{\econj}{\end{conj}}
\newcommand{\bconstr}{\begin{constr}}
	\newcommand{\econstr}{\end{constr}}
\newcommand{\bpf}{\begin{proof}}
	\newcommand{\epf}{\end{proof}}
\newcommand{\bprf}{{\em Proof: }}
\newcommand{\eprf}{\hfill $\Box$}

\begin{document}
	\title{Occupational Network Structure and Vector Assortativity for illustrating patterns of social mobility}
	\author{Vinay Reddy Venumuddala\\ Doctoral Student, Public Policy\\ Indian Institute of Management Bangalore}
	\date{}
	\maketitle
	
\begin{abstract}
	In this study we arrive at a closed form expression for measuring vector assortativity in networks motivated by our use-case which is to observe patterns of social mobility in a society. Based on existing works on social mobility within economics literature, and social reproduction within sociology literature, we  motivate the construction of an occupational network structure to observe mobility patterns. Basing on existing literature, over this structure, we define mobility as assortativity of occupations attributed by the representation of categories such as gender, geography or social groups. We compare the results from our vector assortativity measure and averaged scalar assortativity in the Indian context, relying on NSSO 68th round on employment and unemployment. Our findings indicate that the trends indicated by our vector assortativity measure is very similar to what is indicated by the averaged scalar assortativity index. We discuss some implications of this work and suggest future directions. 
\end{abstract}

\section{Introduction}
In this study, we devise a framework to depict patterns of social mobility with occupations as units of analysis, where we define social mobility as the opposite of durable inequality introduced by Tilly (1998). In his study, Tilly argues that, insofar as the structuration of occupational boundaries within organizations happen along categorical lines, categorical inequalities remain durable. A category for instance can be defined on the basis of gender, religion, or social group, and categorical groups correspond to groups within a given category. In this work, we frame an occupational network structure, with relation between occupations defined by other forces of structuration such as the industry and educational requirements. We define strength of connection between occupations as the similarity of occupations along these non-categorical factors. We also define categorical attributes for each occupation as  a vector capturing the representation of distinct categorical groups in that occupation, relative to their representation in the overall work force. We then devise a vector assortativity measure to depict the stratification of occupational network along categorical lines and proxy it for categorical inequality following Tilly (1998). Observing  this assortativity measure over consecutive birth cohorts, allows us to illustrate the patterns of social mobility given a particular category. Our findings indicate that assortativity along sector is more or less stagnant, along gender it's falling in the recent cohorts, but for social groups it has been consistently increasing with slight stagnation observed in recent cohorts. Further we believe that looking at the changing contribution to the overall assortativity by different industry-education combinations (set of non-categorical structuration forces) through time, can allows us to comment on those structuration forces which are amicable to social mobility and those which perhaps are not.   

\section{Social Mobility and Reproduction}
Studies on social mobility often characterize its measurement as a relative or absolute improvement in socio-economic status of individuals within any society, seen either inter-generationally or intra-generationally \citep{narayan_fair_2018}. Here the socio-economic status of individuals are indicated either by income \citep{becker_equilibrium_1979,black_recent_2010,chetty_where_2014,solon_intergenerational_1999}, or education \citep{asher_intergenerational_2018,azam_like_2015}, or class positions determined by employment relations and defined as aggregate occupational groupings \citep{erikson_intergenerational_2002}. The one commonality among these studies is that the measurement of social mobility is pegged at a macro or societal level. Each of these studies also base their analysis on plausible pre-suppositions and justifications about why they had to choose any particular indicator as an individual's socio-economic status and how does it help in the measurement of social mobility. Further these ontological pre-suppositions also help researchers to provide plausible explanations for the processes/mechanisms that cause variation in social mobility across societies. \\

For instance, rational choice based social mobility models within economics literature rely on individual attributes, both ascriptive and non-ascriptive, to theorize the processes/mechanisms underlying mobility as well as to explain the variation in mobility across societies. For example, the seminal rational choice based theoretical model on intergenerational mobility proposed by \cite{becker_equilibrium_1979}, describe these processes through a dependence between parent's earnings and child's earnings. Such dependence according to them is mediated by factors such as human and non-human capital investments of parents on their children, individual `endowments' determined by ascriptive characteristics such as ethnicity, family connections, and so on. Much of the literature on social mobility within economics rely on empirically evaluating the relative influence of each of these processes/mechanisms across different societies \citep{black_recent_2010,solon_intergenerational_1999}.\\ 

Within sociology literature pre-suppositions are often made about the structure of society, in particular in terms of the class positions that constitute it. The class schema proposed by \cite{erikson_intergenerational_2002} defined using `employment status and occupation as indicator of employment relations' has motivated several empirical works on social mobility within this strand of literature \citep{azam_intergenerational_2015,iversen_rags_2016,motiram_how_2012}. In so far as mechanisms underlying social mobility are concerned, many macro sociological theories, shed light instead into the mechanisms underlying reproduction of social inequalities \citep{,tilly_durable_1998}. \cite{bourdieu_distinction_2013}, for example, illustrates the mechanism underlying social reproduction through the notion of habitus generating action. According to Bourdieu, the processes associated with social stratification and its reproduction can be depicted within a field or a social space in which individuals or groups are closely linked to different class/social positions. Each such position is bound up with systems of dispositions called habitus that is bound up with a particular set of cultural tastes or capital. This habitus in turn, dictates the practices (or actions) of individuals occupying their respective class positions. These individual actions, affected by a misrecognition of historically contingent social relations within the field, tend to perpetuate the very influence (and location in the hierarchy) of a class position ad-infinitium, reproducing social inequality and the observed patterns of social stratification with time \citep{burawoy_making_2018,riley_bourdieus_2017}. For \cite{tilly_durable_1998}, the mechanisms underlying reproduction of social or categorical  inequalities predominantly situate within organizations in any society. \cite{tilly_durable_1998} proposes mechanisms that contribute to installation of the widely recognized categories (such as gender, religion, caste, ethnicity and so on) internally, their maintenance, and their percolation into many organizational forms within society. By explicitly focusing on the persistence of structures that influence world of work within and across organizations, \cite{tilly_durable_1998} provides explanations for the mechanisms that underlie durable categorical inequalities.\\

Empirical works that test much of the macro sociological theories, do so by illustrating the stability of stratification or class structures within societies by making plausible assumptions about the basis of stratification. \cite{bourdieu_distinction_2013} for example, treat cultural capital of individuals as constituting the basis for stratification. Empirically highlighting an association of cultural tastes of individuals with the occupations they are situated in, \cite{bourdieu_distinction_2013} plots occupations onto social space and depict a hierarchy of occupations or the stratification structure in terms of differences in their cultural capital. Recent works within CAMSIS (Cambridge Social Interaction and Stratification)  tradition \citep{bottero_stratification_2004,griffiths_dimensions_2012,lambert_social_2018}, base their stratification structure on social relations between individuals across occupations to determine whether occupations are socially close or distant in a social space. They empirically map occupations onto a two-dimensional space or as a network in order to identify emergent classes or strata based on closely spaced occupations. More recent work by \cite{toubol_mapping_2017} assume occupational mobility as the basis for class formation based on the works of Max Weber. They map occupations onto a network with edge weights determined by the extent of intra-generational mobility between any two occupations, to explore class formations based on clustering patterns within the constructed occupational network. In all these studies occupations are treated as the units of analysis while observing stratification or class structure.

\section{Durable Inequality and Institutionalization of categorical pairs}
\begin{quote}
	\textit{``Again, the founder of a small manufacturing firm, following models already established in the trade, divides the firm’s work into clusters of jobs viewed as distinct in character and qualifications and then recruits workers for those jobs within well-marked categories. As turnover occurs and the firm expands, established workers pass word of available jobs among friends and relatives, collaborating with and supporting them once they join the work force. Those new workers therefore prove more reliable and effective than others hired off the street, and all concerned come to associate job with category, so much so that owner and workers come to believe in the superior fitness of that category's members for the particular line of work.''} \citep{tilly_durable_1998}
\end{quote}

Charles Tilly in his seminal work the `Durable Inequality' \citep{tilly_durable_1998}, proposes mechanisms that operate within and across organizations and which are at the root of persistent inequalities along categories such as gender, race, caste, ethnicity and so on. For him, causes for social inequality and its reproduction within society can be understood in terms of mechanisms that sustain inequalities along categorical groups. The central argument of his thesis is that: ``Large, significant inequalities in advantages among human beings correspond mainly to categorical differences such as black/white, male/female, citizen/foreigner, or Muslim/Jew rather than to individual differences in attributes, propensities, or performances.'' \citep{tilly_durable_1998}. According to him, `durable inequality depends heavily on institutionalization of categorical pairs', and this occurs when organizations at large, match individuals from external unequal categories with internal work roles for the purpose of efficiency and maintenance. Four mechanisms are key to this matching of unequal categorical structures internally. First, the installation of categorical pairs occurs when people commanding resources at the helm of organizations although draw returns from the work of others, they nevertheless exclude others from the full value added by their effort. This is the mechanism of exploitation. Second, when individuals or groups belong to a `categorically bounded network', any resources acquired by such members are supported and often enhanced by the `network’s modus operandi'. This is the mechanism of opportunity hoarding. Third, the mechanism of emulation, happens when established organizational models are copied or replicated by many organizations across the society. Fourth, the elaboration of daily routines within organizations across its internal work boundaries, often happens on the basis of the unequal categorical structures, which is the mechanism of adaptation. While exploitation and opportunity hoarding facilitate the installation of categorical boundaries into organizations, emulation and adaptation generalize their influence across society \citep{tilly_durable_1998}.

\begin{quote}
``\textit{The notion of upward mobility contains the idea that it represents the triumph of individual achievement over structural constraints}'' \citep{bottero_stratification_2004}.
\end{quote}
 So long as categorical boundaries continue to match work boundaries within and across organizations, individuals can also be constrained by such unequal categorical structures within the world of work. Social Mobility through the lens of Tilly, is therefore possible only when the substantive work boundaries become independent of the categorical boundaries. According to \cite{tilly_durable_1998}, introduction of organizational forms and with work structures that foreclose possibility of categorical matching, is the only way out for overcoming `durable inequality'. According to him, ``reduction or intensification of racist, sexist, or xenophobic attitudes will have relatively little impact on durable inequality, whereas the introduction of certain new organizational forms - for example, installing different categories or changing the relation between categories and rewards - will have great impact'' (pg. 19). In essence, the structure of social inequality and mobility, is constituted by persistent inequalities along categories such as gender, caste, religion and so on, that are institutionalized within organizations. And the mechanisms that persist such inequalities within society are also the mechanisms that maintain boundaries between work roles within and across organizations by matching them with categorical boundaries. Approximately equating work roles with occupations, in so far as the occupational boundaries continue to match with categorical boundaries in any society, it indicates the persistent influence of the aforementioned inequality reproducing mechanisms. 

\section{Other forces of structuration acting upon organizations}
\begin{quote}
	Institutions are the resilient social structures, that comprise ``\textit{regulative, normative, and cultural-cognitive elements that, together with associated activities and resources, provide stability and meaning to social life}'' \citep{scott_institutions_2013}.
\end{quote}

In describing the mechanisms underlying persistence of categorical inequalities, \cite{tilly_durable_1998} extensively relies on the stability of organizational forms and the institutionalization of categorical pairs within them. This institutionalization occurs through scripts, practices, established hierarchies, and other forms of networked relationships between work roles, that structure the actions of incoming actors within organizations in such a way that it reinforces the internal work boundaries along categorical lines. \\

However, it is hard to deny that several other forces of structuration do exist, which act upon organizations and alter their forms through time. These forces do not necessarily depend upon institutionalized practices, scirpts, and relationships built around categorical pairs. Instead, they depend upon the particular organizational field in which organizations are located and the corresponding rules, norms, practices that govern such fields. According to \cite{dimaggio_iron_1983} organizations that produce similar services or products, or say, belong to a particular industry, while in aggregate constitute an organizational field, they are simultaneously subject to similar institutional pressures that make them isomorphic to one another. In their seminal work on `institutional isomorphism', which is at the foundation of institutional theory literature, \cite{dimaggio_iron_1983} highlight three important mechanisms that capture the mechanisms or the forces of structuration which influence organizations to become similar to one another in an organizational field.  These mechanisms are broadly effected by the state (coercive), uncertainties of the market (mimetic), and lastly the professions (normative). In so far as the role of state is concerned, the mechanism of coercive isomorphism explains for example, how and why organizational changes are a response to `government mandates', or other `legal and technical requirements of the state'. The mechanism of mimetic isomorphism explains as to how and why, in response to uncertainties of the market, organizations tend to morph themselves similarly with other organizations that are perceived to be legitimate or successful within their field. Lastly, professionalization makes organizations similar within a field through the mechanism of normative isomorphism. Following \cite{sarfatti_larson_rise_1977}, \cite{dimaggio_iron_1983} define the idea of professionalization as the ``collective struggle of members of an occupation to define the conditions and methods of their work, to control, `the production of producers' (\cite{sarfatti_larson_rise_1977}: 49-52), and to establish a cognitive base and legitimation for their occupational autonomy''. Formal education, and professional networks spanning organizations, according to them are the two important aspects of professionalization that contribute to homogeneity of organizational structures within a given field and the consequent variation across fields.\\

From the above discussion we note that the changes in political and economic structure of the state, the market environment, and changes around professionalization, together can alter the organizational forms within a given field largely independent of the institutionalization of categorical pairs. Therefore, such changes in organizational forms could also contain in them the seeds of social mobility \citep{tilly_durable_1998}, where other forces of structuration discussed above, can overpower the institutionalization of categorical pairs. In the subsequent section, we propose occupational network structure as a framework that allows us to look at these two kinds of forces separately, and 1. Identify the patterns of social mobility and 2. Look out for the probable mechanisms or forces of structuration that could have contributed to social mobility. 



\section{Occupational Network Structure (ONS)}
According to \cite{tilly_durable_1998}, matching of external categories (such as religion, gender, language, caste and so on) onto internal work roles within organizations is critical to the reproduction of categorical inequalities. Insofar as work roles or occupations are structured and therefore segregated or stratified along categorical lines, it is indicative that the mechanisms of durable inequality are at play. That is, in such case, the forces of structuration operating along categorical pairs, also cast their influence over the proximity or distance between occupations in the substantive world of work. In the earlier section we have seen that organizations are structured by the institutional influences of the corresponding organizational field. These institutional influences therefore also tend to structure the occupations that comprise such organizations. Further we have also seen that occupations by themselves are structured by forces beyond organizational field such as the formal education of its occupants and the professional networks that undergrid such occupations \citep{sarfatti_larson_rise_1977}. Infact \cite{grusky_decomposition_2001} emphasize that the substantive forces of structuration happen locally at the level of disaggregated occupations, rather than aggregate classes. Such a structuration, according to them, could manifest as `tangible and symbolic control over the supply of labour' in the form of stipulated educational requirements, instituted licensing systems, organized unions, occupational associations and so on \citep{grusky_decomposition_2001}.\\

Above we have seen above, there are two sets of structuration forces that are at play operating over occupations within organizations. One as per \cite{tilly_durable_1998}, that says institutionalized categorical pairs determine occupational boundaries. Second, which say that occupations within organizations are structured by the institutional forces operating on the corresponding organizational fields and also by the stipulated educational requirements, licensing systems, unions, associations and so on \citep{dimaggio_iron_1983,grusky_decomposition_2001}. In order to evaluate social mobility we therefore attempt to look at the operation of these different sets of structuration forces over occupations, separately in order to check for social mobility patterns. We conceive of an occupational network structure that is constructed based on the forces of stratification that are non-categorical in nature, and look for the extent to which such a network is also stratified along attributes defined by categorical representations in each occupation. We describe this network construction below, and discuss the mathematical aspects associated with measurement of social mobility and mechanisms in the subsequent section. 

\subsection{ONS construction}
For us, given a category (such as gender), each occupation is defined by 1. Representation  of individuals across groups within such category (such as male/female in case of gender), relative to their representation in the overall workforce, 2. Distribution of workforce in this occupation across different industries, and 3. Distribution of workforce in this occupation across different education levels. We proxy (2) for the structuration forces operating on an occupation that come from its association with multiple organizational fields, and (3) for the stipulated educational requirements specific to an occupation.  Note that, owing to data availability we limit to only (2) and (3) for proxying the structuration forces that are specific to an occupation, however in reality one could include other factors such as licencing requirements, associations and so on. We consider (1) to be the label attributes of occupations that are important while measuring stratification along categorical lines using graph assortativity measures as we shall see in subsequent sections.\\

Before going into the details one could ask at this point, as to why we do not look at each occupation as a separate entity and depict the structure of world of work as constituted by bundle of independent occupations? Or, one might also ask, could you not impose a hierarchy of some sort to the occupations \footnote{For example, bundling occupations along hierarchically situated class locations following established class schemas such as the one proposed by \cite{erikson_intergenerational_2002}, or rank occupations on the basis of average education or average income of the individuals.}? We acknowledge that both these ways of looking at the occupational structure are plausible, where in the former case there is no notion of distance between occupations, and in the latter, there is a clear-cut hierarchy imposed to define the occupational structure. However, given our conceptualization about the structure of the world of work we find occupational network as lying somewhere in between these two extremes. In addition it serves as a framework to observe social mobility patterns and simultaneously look for changes in the occupational network that could possibly have answers for mechanisms or structuration forces that contribute to social mobility.\\

\begin{defn}
	\textbf{Occupational Network: }Let $E = \{E_1,E_2,..,E_m\}$ be set of education levels, $I = \{I_1,I_2,...,I_n\}$ be set of industry sectors. Consider a distribution defined over support $S := E\times I$. We denote the non-categorical structuration forces operating on an occupation at a given point in time (or for a given birth cohort) by the distribution of individuals associated with it over the support $S$. Now the distance between occupation $O_i$ and $O_j$ can be given by the distributional distance $D$ between $O_i$ and $O_j$ defined over the support $S$. This means, if two occupations are closer that indicates that the non-categorical structuration forces operating on them are similar. We use total variation distance as our measure of distributional distance as it is always bounded in the range $[0,1]$, with $0$ indicating exact same distributions, while $1$ indicates completely disjoint distributions. We transform distributional distance into edge weight, where edge weight is given by $w_{i,j} = 1 - D(O_i,O_j)$. After constructing an adjacency matrix with these edge-weights, we remove the edge-weights that connect the same occupations to avoid self-loops, and subsequently normalize its values by the sum total of all the remaining edge-weights. Our resultant network is therefore an undirected weighted network with weights given by corresponding values in the normalized adjacency matrix $A$. 
\end{defn}

In the following section, we first introduce scalar assortativity measure defined over undirected weighted graphs, where node labels are given by scalar real valued attributes. In our case, however, with nodes as occupations, we have its vector attributes denoted by the representation of individuals across groups within any category relative to overall workforce (1). We therefore extend the scalar assortativity measure to the case of vector attributes in a graph, so as to depict the extent of assortativity along categorical lines on a given occupational network. 

\section{Network Assortativity Measures for depicting mobility patterns}
Here we first discuss the assortativity measure over graphs with nodes having scalar attributes. \cite{newman_mixing_2003} defines this measure over scalar attribute graphs as Pearson's correlation coefficient across edges. We first adjust this correlation coefficient to weighted graphs, and make explicit, the notion of attributes of nodes on an edge as random variables, and edge-weights as corresponding probabilities, following \cite{peel_multiscale_2018}. Applying similar intuition, we then extend the scalar assortativity measure to vector attributes by replacing Pearson's correlation coefficient with distance correlation \citep{lyons_distance_2013,szekely_measuring_2007-1}. We trade-off the interpretation of linear independence as offered by assortativity measure based on Pearson's correlation coefficient with the interpretation of independence. Since our attribute space is comprised of vectors of a chosen dimension, we consider this assumption as a plausible one to make, given our problem.

\subsection{Scalar assortativity Measure}
Assortativity defines the property of a network or a graph where nodes with similar attributes have a tendency to be strongly connected than those with dissimilar attributes \citep{newman_mixing_2003}. In the context of scalar node attributes (such as age) \cite{newman_mixing_2003} indicates that assortative mixing over a social network could suggest stratification of society along such attributes. He defines assortativity measure as Pearson's correlation coefficient measured between attributes of adjacent nodes in the network. In the following proposition we adjust this measure to the case of a weighted graph following \cite{peel_multiscale_2018}. We also explicate the assumptions that underpin this approach of measurement, which as we will show subsequently, will help us to extend this measure to graphs with nodes having vector attributes. 

\bprop
	Consider an undirected weighted graph $G = (V, E)$ having $n$ nodes and $m$ edges. Every edge in the graph is defined by pair of nodes $(i,j)$, with strength of the connection given by a weight $w_{ij}$, and attributes of the nodes given by $x_i$ and $x_j$ such that $x_i,x_j \in \mathbb{R}$. Assortativity measure $r$ on this network following \cite{newman_mixing_2003} and \cite{peel_multiscale_2018} is given by 
	\begin{eqnarray*}
	r = \frac{\sum_{ij}A_{ij}(x_i -\bar{x}) (x_j - \bar{x})}{\sum_i k_i (x_i - \bar{x})^2}
	\end{eqnarray*}
	where $A$ denotes normalized adjacency matrix such that $A_{ij} = \begin{cases}
	\frac{w_{ij}}{\big( \sum_{i\le j} w_{ij} \big)}, \ if \ i = j\\
	\frac{w_{ij}}{2\big(\sum_{i\le j} w_{ij} \big)}, \ if \ i \ne j\\
	\end{cases}$, $k_i = \sum_j A_{ij}$, and $\bar{x} = \sum_i x_i k_i$.
\eprop
\bprf
Consider $X$ and $Y$ to be random variables denoting the attributes of start and end nodes corresponding to any randomly chosen edge in the graph (Note that every undirected edge between two different nodes is treated as two directed edges). $X$ and $Y$ follow the same distribution with support defined over all the scalar attribute values over the undirected graph. Joint distribution of $X,Y$, is defined in terms of edge-weights as follows,

\begin{align*}
P(X=x,Y=y) = \sum\limits_{\substack{i,j \in V\\ x_i = x, x_j = y}} A_{ij}
\end{align*}

Assortativity index according to \cite{newman_mixing_2003} is given by Pearson's correlation coefficient,
\begin{eqnarray*}
r &=& \frac{E[(X-\mu_x)(Y-\mu_y)]}{\big(E[(X-\mu_x)^2] E[(Y-\mu_y)^2]\big)^{1/2}}\\
&=& \frac{E[(X-\mu)(Y-\mu)]}{E[(X-\mu)^2]} \text{  (Since $X$ and $Y$ have identical marginal distributions with mean $\mu$)}\\
&=& \frac{\sum_{x,y} (x-\mu) (y - \mu) P(X = x, Y= y)}{\sum_{x} (x-\mu)^2 P(X = x)}\\
&=& \frac{\sum_{ij}A_{ij}(x_i -\mu) (x_j - \mu)}{\sum_i k_i (x_i - \mu)^2} \ \ \ \left(\text{ as }  P(X = x, Y= y) = \sum\limits_{\substack{i,j \in V\\ x_i = x, x_j = y}} A_{ij}, \ \ P(X = x) = \sum\limits_{y} P(X=x, Y=y)\right)\\
&& \text{Which is same as Equation (B2) in \cite{peel_multiscale_2018}}
\end{eqnarray*}
\eprf

\subsection{Vector Assortativity Measure}
In the above formulation of assortativity index, the probability $P(X=x,Y=y)$ indicates the fraction of edges that connect nodes having the attribute $x$ with nodes having attribute $y$. $A_{ij}$ on the other hand indicates the strength of the connection between two nodes $i$ and $j$. Our idea behind illustrating the above formulation is to bring out a distributional assumption that the strength of an edge is proportional to the probability of selecting adjacent nodes (as the start and end nodes) constituting that edge. A higher assortativity therefore indicates that strongly connected nodes, whose consequent edges are also more probable of selection, are also close in terms of their corresponding node attributes. In our case where the network is constituted by occupations and their interconnections, the aforementioned index reflects the stratification of occupations along any scalar real-valued attribute defined over each of the occupations. A higher assortativity indicates that occupations are largely stratified along this attribute. \\

However, in so far as our study is concerned, we are interested in categorical attributes of occupations which can be multi-dimensional. Since we are interested in stratification of occupations along categorical attributes, we associate a vector made up of representations of different groups within a given category to each occupation relative to their corresponding representation in the total workforce. For example, if we are interested in stratification based on sector, which is a category constituted by two groups rural and urban, then the vector constituted by representations of rural and urban workforce in given occupation, relative to the rural and urban workforce in the total population, defines the corresponding occupation's attribute. Although in categories defined by only two groups, representation of any one category can be treated as scalar real-valued attribute to compute assortativity index like above, but for those defined by more than two groups (such as language, religion, caste, ethnicity and so on) we will have to accommodate vector attributes as well. We can also treat parent's occupations as categorical groups, which is usually the case in contingency table based mobility measurement approaches.  \\

This motivates us to define a vector assortativity index to measure stratification of occupations along categorical attributes, which captures the essence of categorical inequality in Tilly's terms. Observing how inequalities along various categories such as religion, gender, caste and so on, are changing over time, will tell us whether or not the substantive work boundaries are becoming independent of given categorical boundaries. Higher the assortativity, higher is the extent of categorical inequality, and persistence of a higher assortativity through time relates to absence of social mobility. In other words, vector assortativity helps us infer about whether the changes in the occupational network structure through time  is reflects social mobility or the reproduction of categorical inequalities.  \\

To construct a vector assortativity index, we build on the notion of linear dependence (or correlation) of real-valued scalar random variables (denoting node attributes), and extend it to multi-dimensional real valued random vectors (also denoting node attributes). Since linear dependence of random vectors has little meaning (as signified by the usage of Pearson's correlation coefficient), we instead consider independence of random vectors. We therefore replace Pearson's correlation coefficient with distance correlation \citep{szekely_measuring_2007-1,lyons_distance_2013} between two random vectors to extend the above proposition to vector attributes of nodes. We illustrate it in the following proposition.

\bprop
Consider an undirected weighted graph $G = (V, E)$ having $n$ nodes and $m$ edges. Every edge in the graph is defined by pair of nodes $(i,j)$, with strength of the connection given by a weight $w_{ij}$, and attributes of the nodes given by $x_i$ and $x_j$ such that $x_i,x_j \in \mathbb{R}^d$. Vector assortativity measure $r$ on this network is given by a distance correlation measure, following \cite{szekely_measuring_2007-1}, and \cite{lyons_distance_2013}, which is the square root of
\begin{eqnarray*}
	r^2 = \frac{f_1}{f_2}
\end{eqnarray*}
Where $f_1$ is given by the following expression
\begin{eqnarray*}
\sum_{i',j'} \sum_{i,j} A_{ij} A_{i'j'} d(x_i,x_i') d(x_j, x_j') - 2 \sum_{i,j} A_{ij} \left( \sum_{i'} A_{i'.} d(x_i,x_i') \right) \left( \sum_{j'} A_{.j'} d(x_j,x_j')\right) + \left(\sum_{i,i'} A_{i.}A_{i'.} d(x_i,x_i') \right)^2
\end{eqnarray*}
and $f_2$ is given by the following expression
\begin{eqnarray*}
\sum_{i',j'} A_{i.}A_{i'.}(d(x_i,x_i'))^2 - 2 \sum_{i} A_{i.}\left(\sum_{i'} A_{i'.} d(x_i,x_i') \right)^2 + \left(\sum_{i,i'} A_{i.}A_{i'.}d(x_i,x_i')\right)^2
\end{eqnarray*}

 $A$ here denotes normalized adjacency matrix such that $A_{ij} = \begin{cases}
\frac{w_{ij}}{\big( \sum_{i\le j} w_{ij} \big)}, \ if \ i = j\\
\frac{w_{ij}}{2\big(\sum_{i\le j} w_{ij} \big)}, \ if \ i \ne j\\
\end{cases}$, $A_{i.} = \sum_j A_{ij}$, and $d(x_i,x_j) = ||x_i - x_j||$, where $||.||$ denotes Euclidean norm.
\eprop
\bprf
Consider $X$ and $Y$ to be random vectors denoting the vector attributes of start and end nodes corresponding to a randomly selected edge from the network (Note that every undirected edge between two different nodes is treated as two directed edges). $X$ and $Y$ follow the same distribution with support defined over all possible vector attributes over the undirected graph. Joint distribution of $X,Y$, is defined in terms of edge-weights as follows,

\begin{align*}
P(X=x,Y=y) = \sum\limits_{\substack{i,j \in V\\ x_i = x, x_j = y}} A_{ij}
\end{align*}

Since we have vector attributes, instead of Pearson's correlation coefficient here we consider population distance correlation which was first defined by \cite{szekely_measuring_2007-1}. We follow an equivalent definition of distance covariance given by \cite{lyons_distance_2013}, in order to arrive at a simplified form given that our population distribution is completely determined by the normalized adjacency matrix $A$.\\

Consider $(X',Y')$ be independent and identically distributed copies of $(X,Y)$. Then following \cite{lyons_distance_2013}, the distance covariance and variance are given by,
\begin{eqnarray*}
 dCov^2(X,Y) &:=& E[d_\mu(X,X') d_\nu(Y,Y')],\\
 dCov^2(X,X) &:=& E[d_\mu(X,X')^2],\\
\text{where, } d_\mu(X,X') &=& d(X,X') - a_\mu(X) - a_\mu(X') + D(\mu) \text{ and,}\\
 \ d_\nu(Y,Y') &=& d(Y,Y') - a_\nu(Y) - a_\nu(Y') + D(\nu)\\
\text{and }  d(X,X') &=& ||X-X'||, \ a_\mu(X) := E_{X'}[||X-X'||] \\
\text{and }  D(\nu) &=& E[||X-X'||]
\end{eqnarray*}
$\mu$ and $\nu$ represent the distributions followed by $X$ and $Y$ respectively. However in our case $\mu = \nu$.

We define vector assortativity index as distance correlation, which following \cite{szekely_measuring_2007-1}, and \cite{lyons_distance_2013} is given by square root of,
\begin{eqnarray*}
r^2 = \frac{dCov^2(X,Y)}{(dCov^2(X,X) \times dCov^2(Y,Y))^{\frac{1}{2}}}
\end{eqnarray*}
We first simplify the numerator of the above expression, 
\begin{eqnarray*}
dCov^2(X,Y) &=& E[d_\mu(X,X') d_\nu(Y,Y')]\\
&= & \sum_{x',y'} \left(\sum_{x,y} d_\nu(X,X') d_\nu(Y,Y') P(X=x,Y=y) \right)P(X'=x',Y'=y')
\end{eqnarray*}
\text{Expanding the expression, } $d_\nu(X,X') d_\nu(Y,Y') = \\ \left(d(x,x') - a_\mu(x) - a_\mu(x') + D(\mu)\right) \left(d(y,y') - a_\nu(y) - a_\nu(y') + D(\nu)\right)$, we have the following sets of terms.
\begin{enumerate}
	\item $d(x,x') d(y,y')$
	\item $-a_\nu(y)d(x,x'),-a_\mu(x)d(y,y'), a_\mu(x)a_\nu(y),-a_\mu(x')d(y,y'),-a_\nu(y')d(x,x'), a_\mu(x')a_\nu(y')$
	\item $D(\mu) (d(y,y') - a_\nu (y)),  D(\nu) (d(x,x') - a_\mu (x))$
	\item $D(\mu)D(\nu), -D(\mu)a_\nu (y'), -a_\mu(x') D(\nu), a_\mu(x')a_\nu(y), a_\mu(x)a_\nu(y')$
\end{enumerate}
Let us consider expanding the first term in (2) over the summations defining covariance 
\begin{eqnarray*}
&&\sum_{x',y'} \left(\sum_{x,y} -a_\nu(y)d(x,x') P(X=x,Y=y) \right)P(X'=x',Y'=y')\\
&& \ \ \ = \sum_{x,y} \left(\sum_{x',y'} -a_\nu(y)d(x,x') P(X'=x',Y'=y') \right) P(X=x,Y=y)\\
&& \ \ \ = \sum_{x,y} -a_\nu(y)\left(\sum_{x'} d(x,x') P(X'=x') \right) P(X=x,Y=y)\\
&& \ \ \ = \sum_{x,y} -a_\nu(y)a_\mu(x) P(X=x,Y=y)\\
\end{eqnarray*}
Since distributions $\mu$ and $\nu$ are identical and by symmetry it follows that all the terms in (2) without the sign end up with same value. Therefore all the terms in (2) together simplify as $-2\sum_{x,y} a_\nu(y)a_\mu(x) P(X=x,Y=y)$ \\
Now consider the first term in (3), we have
\begin{eqnarray*}
&&\sum_{x',y'} \left(\sum_{x,y} D(\mu) (d(y,y') - a_\nu (y)) P(X=x,Y=y) \right)P(X'=x',Y'=y')\\
&& \ \ \ = D(\mu) \sum_{x',y'} \left(\sum_{x,y} (d(y,y') - a_\nu (y)) P(X=x,Y=y) \right)P(X'=x',Y'=y')\\
&& \ \ \ = D(\mu) \sum_{x,y} \left(\sum_{x',y'} (- a_\nu (y)) P(X'=x',Y'=y') \right)P(X=x,Y=y) + \\
&& \ \ \ \ \ \ D(\mu) \sum_{x,y} \left(\sum_{x',y'} d(y,y') P(X'=x',Y'=y') \right)P(X=x,Y=y)\\
&& \ \ \ = - D(\mu) \sum_{x,y} a_\nu (y) P(X=x,Y=y) +  D(\mu) \sum_{x,y} a_\nu (y) P(X=x,Y=y)\\
&& \ \ \ = 0
\end{eqnarray*}
Similarly the other term in (3) also vanishes\\

Now consider the third and fourth terms together in (4)\\
\begin{eqnarray*}
&&\sum_{x',y'} \left(\sum_{x,y} \left[ -a_\mu(x') D(\nu) + a_\mu(x')a_\nu(y) \right] P(X=x,Y=y) \right)P(X'=x',Y'=y')\\
&& \ \ \ = \sum_{x',y'} -a_\mu(x') \left(\sum_{x,y}  \left[ D(\nu) - a_\nu(y) \right] P(X=x,Y=y) \right)P(X'=x',Y'=y')\\
&& \ \ \ = \sum_{x',y'} -a_\mu(x') \left( D(\nu) - \sum_{x,y}  a_\nu(y)  P(X=x,Y=y) \right)P(X'=x',Y'=y')\\
&& \ \ \ = \sum_{x',y'} -a_\mu(x') \left( D(\nu) - \sum_{y}  a_\nu(y)  P(Y=y) \right)P(X'=x',Y'=y')\\
&& \ \ \ = 0 \text{ , since } \sum_{y}  a_\nu(y)  P(Y=y) = D(\nu) \\
\end{eqnarray*}
Similarly terms two and five in (4) also cancel each other out, and only the first term remains. Therefore expression for $dCov^2(X,Y)$ is eventually simplified as
\begin{eqnarray*}
dCov^2(X,Y) &=& E[d_\mu(X,X') d_\nu(Y,Y')]\\
&= & \left(\sum_{x,y} d_\nu(X,X') d_\nu(Y,Y') P(X=x,Y=y) \right)P(X'=x',Y'=y')\\
&= & \sum_{x',y'} \left(\sum_{x,y} d(x,x') d(y,y') P(X=x,Y=y) \right) P(X'=x',Y'=y')\\
&& \ \ \ \ \ \ - 2 \sum_{x,y} a_\nu(y)a_\mu(x) P(X=x,Y=y) + D(\mu)^2 \text{ , since $\nu$, and $\mu$ have same support.} \\
&& \text{Now using } P(X=x,Y=y) = \sum\limits_{\substack{i,j \in V\\ x_i = x, x_j = y}} A_{ij}, \text{ we get,}\\
dCov^2(X,Y) &=& \sum_{i',j'} \sum_{i,j} A_{ij} A_{i'j'} d(x_i,x_i') d(x_j, x_j') - 2 \sum_{i,j} A_{ij} \left( \sum_{i'} A_{i'.} d(x_i,x_i') \right) \left( \sum_{j'} A_{.j'} d(x_j,x_j')\right)\\
&& \ \ \ \ \ \  + \left( \sum_{i,i'} A_{i.}A_{i'.} d(x_i.x_i.)\right)^2
\end{eqnarray*}

Similarly, we can simplify the denominator of $r^2$, given by $(dCov^2(X,X) \times dCov^2(Y,Y))^{\frac{1}{2}} =  dCov^2(X,X)$ to the following expression,
\begin{eqnarray*}
\sum_{i',j'} A_{i.}A_{i'.}(d(x_i,x_i'))^2 - 2 \sum_{i} A_{i.}\left(\sum_{i'} A_{i'.} d(x_i,x_i') \right)^2 + \left(\sum_{i,i'} A_{i.}A_{i'.}d(x_i,x_i')\right)^2
\end{eqnarray*}
This is the end of our simplification.
\eprf

\section{Data and Findings}
For the purpose of our study we carry out our analysis using NSS 68th round on employment and unemployment which captures details of individual occupation, industry and education. We recode education along four levels, 1. Below Primary, 2. Below Secondary and above primary, 3. Below Graduation but above secondary, and 4. Above graduation. Industry information is captured in the survey according to NIC-2008 code structure, and we consider this information at the least disaggregated level captured by 20 section codes. Occupation information is captured according to NCO-2004 code structure, and we consider 2-digit codes for the purpose of our study. 
\subsection{Patterns of social mobility}
For a given birth cohort we build network based on our definition of occupational network. From year 1940 to year 1980 we consider consider consecutive and overlapping 10 year birth cohorts considering sliding windows with spacing of one year. Following are the patterns of vector assortativity and averaged scalar assortativity observed across years. We compute averaged scalar assortativity as the average of scalar assortativity computed based on the proportional representation (relative to workforce), for each of the categorical groups within a given category, which is a scalar label attribute for each occupation. We find that the trends observed for each category are similar following either of these measures (See Figures 1 and 2). Either way we find that while assortativity along sector (rural/urban) is more or less stagnant across cohorts, assortativity along gender (male/female) has come down slightly for the recent cohorts, and has been steadily increasing over the years along social group (GEN/OBC/SC/ST) with a slight dip in the recent cohorts. 
\begin{table}[!ht]
	\begin{minipage}{.5\columnwidth}
		\begin{Figure}
			\captionsetup{font=scriptsize}
			\begin{center}
				\includegraphics[width=3.5in]{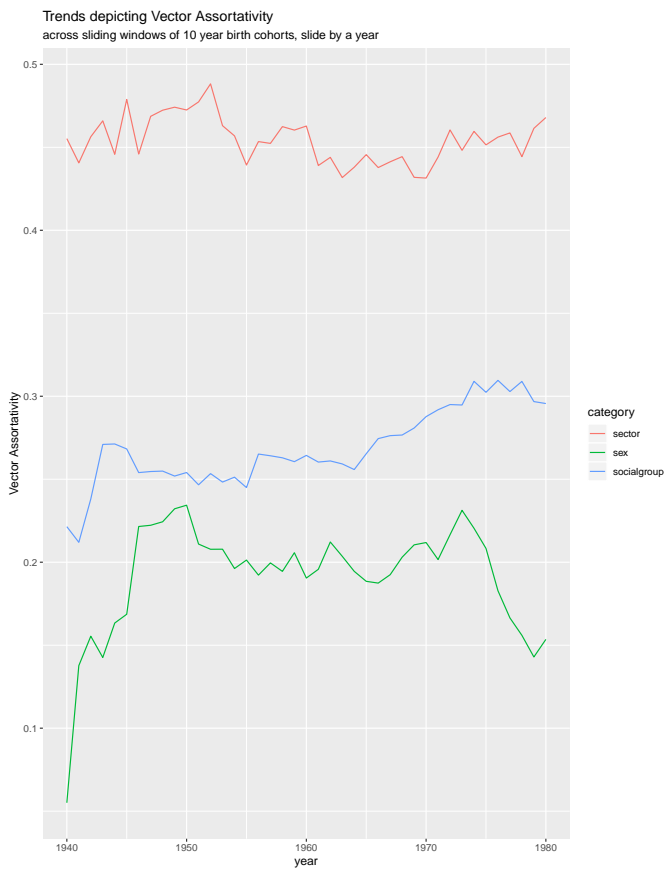}
				\captionof{figure}{Vector Assortativity} 
				\label{fig:VA}
			\end{center}	
		\end{Figure}
	\end{minipage}
	\begin{minipage}{.5\columnwidth}
		\begin{Figure}
			\captionsetup{font=scriptsize}
			\begin{center}
				\includegraphics[width=3.5in]{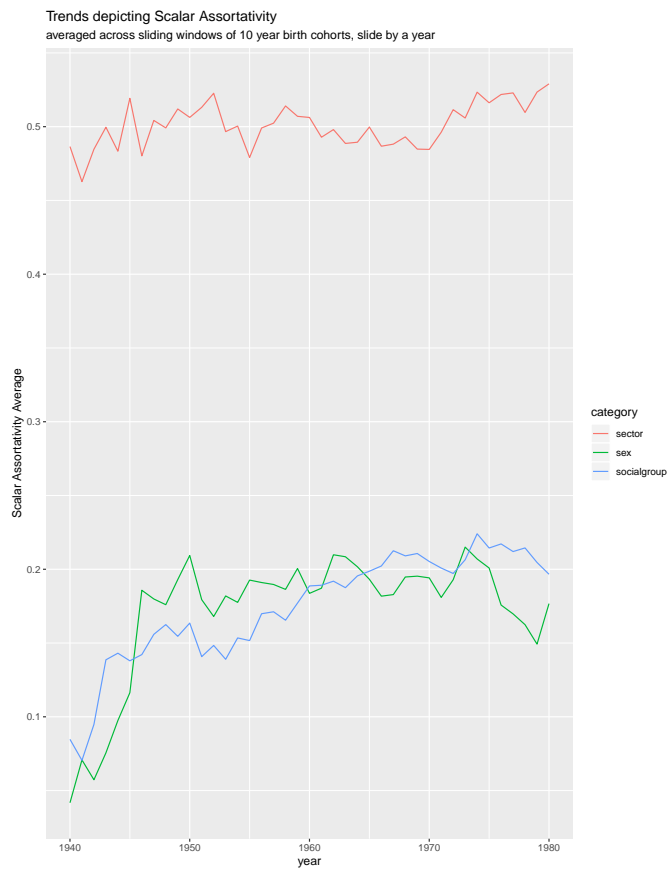}
				\captionof{figure}{Scalar Assortativity} 
				\label{fig:SA}
			\end{center}	
		\end{Figure}
	\end{minipage}
\end{table} 

\section{Implications and Future Work}
In so far as occupations are less assortative along ascriptive categorical attributes, it indicates that the structural constraints determined by such ascriptive characteristics are less important in the world of work.
Within the social mobility literature dealing with intergenerational transition of individuals from one occupational position to another \citep{motiram_how_2012,iversen_rags_2016,azam_intergenerational_2015}, the categorical structures constratining individuals are assumed to be their initial class/occupational positions. These initial class positions are usually indicated by the occupation of their respective parents. Our framework also allows us to define category as constituted by groups of individuals with similar parent occupations. This work addresses key data limitation in developing countries where mobility measurement using income indicators is difficult. Our approach treats occupations as units of analysis, and since occupation information is typically found in most of the sample surveys and even census, it allows for mobility measurement despite such data limitations. \\

Our study also makes a minor contribution to network science literature. Although there are existing works that deal with computing vector assortativity (see \cite{pelechrinis2016va}), here we attempt to provide a closed form expression for computing it based on distance correlation following \cite{szekely_measuring_2007-1} and \cite{lyons_distance_2013}.  \\

\textbf{Future Work}: In order to identify social mobility mechanisms, \cite{tilly_durable_1998} argues that one can find them only when one deep-dives to understand about the forces of structuration that influence work roles or occupations within organizations. Since we considered industry education combinations as the set of non-categorical structuration forces that operate over the occupations within the world of work, one could also attempt to identify or measure the contribution to assortativity by each one of these combinations. 

\bibliographystyle{apalike}
\bibliography{ZoteroLibrary}

\end{document}